\documentclass[conference]{IEEEtran}
\IEEEoverridecommandlockouts

\usepackage{cite}
\usepackage{amsmath,amssymb,amsfonts}

\usepackage{algorithm}

\usepackage{algpseudocode}

\usepackage{amsmath}
\usepackage{amssymb}

\usepackage{graphicx}
\usepackage{textcomp}
\usepackage{xcolor}
\usepackage{booktabs}
\usepackage{url}
\def\BibTeX{{\rm B\kern-.05em{\sc i\kern-.025em b}\kern-.08em
    T\kern-.1667em\lower.7ex\hbox{E}\kern-.125emX}}

\usepackage{draftwatermark}
\SetWatermarkText{ACCEPTED \\MANUSCRIPT}
\SetWatermarkScale{.6}
\SetWatermarkColor[gray]{0.9}

\begin{document}

\title{Software Design Patterns for Latency Measurements with eBPF in SDN Environments\thanks{This manuscript has been accepted for presentation at the IEEE International Symposium on Computers and Communications (ISCC 2026).}
}


\author{
 \IEEEauthorblockN{
 Pedro Lucas M. S. Rosa\IEEEauthorrefmark{1},
 Bruno L. Dalmazo\IEEEauthorrefmark{2},
 Leticia S. Machado\IEEEauthorrefmark{1},
 Jéferson Campos Nobre\IEEEauthorrefmark{1}
 }
 \IEEEauthorblockA{\IEEEauthorrefmark{1}
 Institute of Informatics\\
 Federal University of Rio Grande do Sul, Porto Alegre--RS, Brazil\\
 pedro.sousa@ufgrs.br, leticia.machado@inf.ufrgs.br, jcnobre@inf.ufrgs.br
 }
 \IEEEauthorblockA{\IEEEauthorrefmark{2}
 Computing Science Center\\
 Federal University of Rio Grande, Rio Grande--RS, Brazil\\
 dalmazo@furg.br}
 }

\maketitle

\begin{abstract}
This work proposes, implements, and evaluates a modular architecture based on software design patterns applied to latency measurement in Software-Defined Networks (SDN) using the \textit{Extended Berkeley Packet Filter} (eBPF) and the \textit{Express Data Path} (XDP). Unlike approaches that focus solely on measurement precision, this study examines how design patterns can structure observability systems in an extensible and maintainable way while preserving data-plane efficiency. The proposed architecture applies the \textit{Observer}, \textit{Strategy}, and \textit{Factory Method} patterns to modularize the metric collection and analysis pipeline, allowing dynamic integration of policies and data sinks. The experimental environment was implemented with \textit{Mininet}, \textit{Python} agents, and eBPF/XDP programs, and was evaluated in terms of processing time, overhead, and evolution effort. The results show that the introduction of design patterns show no clear evidence of measurable degradation or performance, maintaining continuous measurements with a minimum of 5\% overhead. The findings suggest that software engineering practices may be applied to the construction of eBPF-based monitoring and observability systems.
\end{abstract}

\begin{IEEEkeywords}
eBPF, XDP, One-Way Delay, Network Observability, Software Design Patterns, Kernel-Level Monitoring.
\end{IEEEkeywords}

\section{Introduction}

The advancement of communication technologies and the growing demand for distributed, latency-sensitive applications have motivated the exploration of new approaches to network monitoring. In particular, the \textit{Extended Berkeley Packet Filter} (eBPF) and the \textit{Express Data Path} (XDP) have emerged as useful tools for kernel-level instrumentation, enabling relatively precise measurements with modest overhead \cite{11325939}. These mechanisms have been applied in different domains, from cloud systems and microservices \cite{levin2020viperprobe,sharma2024ebpf} to software-defined networking (SDN), providing improved visibility of traffic and programmable network functions \cite{castanho2019cadeia}.

Although eBPF/XDP enable highly efficient and precise latency measurements, existing solutions often treat the software architecture of the collection pipeline as a secondary concern. As a result, many approaches prioritize data-plane instrumentation while neglecting architectural aspects such as modularity, extensibility, and controlled evolution, leading to tightly coupled and hard-to-evolve systems. This lack of architectural guidance constitutes a recurring software engineering problem that hinders maintainability, long-term reuse, and experimental reproducibility of eBPF/XDP-based observability solutions~\cite{santos2019aplicaccoes}.

However, most existing studies focus on point solutions for observability without adopting 
software design patterns that promote reuse and modular organization~\cite{abranches2021efficient}. This limitation makes eBPF/XDP-based monitoring solutions harder to maintain and extend, and reduces their experimental reproducibility~\cite{11326005}. 
Thus, the main issue explored in this work is not latency measurement itself, but the lack of structured guidelines to support the development of measurement systems based on eBPF/XDP. We therefore investigate how design patterns can be applied to structure and evaluate latency measurement architectures in a systematic, sustainable, and extensible way. To the best of our knowledge, prior works evaluate eBPF-based latency measurement primarily from a dataplane optimization perspective, without explicitly formalizing the collector architecture using software design patterns.

Recent research demonstrates eBPF efficiency in real-time metric collection, highlighting its precision and scalability \cite{sundberg2023efficient,xie2024analysis,repetto2022monitoring}. Other works address its integration with security and observability mechanisms in containerized and cloud environments \cite{nam2022secure,feng2024enhancing,jin2024enhanced}. Nonetheless, these approaches emphasize dataplaneaspects while neglecting the underlying software design. In the SDN context, studies such as \cite{pacifico2020sistema,carvalho2022bg} reinforce the need for modular architectures that reconcile performance with design clarity, a gap this article aims to fill by formalizing the use of design patterns applied to latency measurement.

Based on this gap, this work proposes a modular and standardized architecture for one-way delay (OWD) measurements in programmable networks, integrating eBPF/XDP with a software layer structured around classic engineering patterns. The architecture comprises three layers: (i) the data plane, implemented in the kernel via XDP; (ii) the user-space control plane, with Python agents; and (iii) a modular HTTP collector structured with the \textit{Observer}, \textit{Strategy}, and \textit{Factory Method} patterns. This integration enables continuous measurements with reasonable accuracy with minimal overhead and supports dynamic reconfiguration of the observability pipeline.

The contributions of this article include: (i) the design of an architectural model that applies software design patterns to latency monitoring using eBPF/XDP, targeting higher-speed dataplane environments; (ii) the implementation and experimental validation of this model in an SDN setting using Mininet; and (iii) a quantitative and qualitative analysis of the impact of modularity and abstraction on extensibility and dataplane performance. The results demonstrate that continuous latency monitoring can be achieved with software modularization while maintaining dataplane overhead below 5\%.

The remainder of this article is organized as follows: Section II presents theoretical foundations and related work; Section III describes the methodology and proposed architecture; Section IV details the experimental environment and results; Section V discusses observations and threats to validity; and Section VI concludes and outlines future directions.

\section{Related Work}

Several recent studies have explored the use of eBPF and XDP for network monitoring and observability of distributed systems. However, most concentrate on specific optimizations or on cloud and microservice scenarios, without directly addressing the application of software design patterns to continuous latency measurements.

\textit{Efficient Continuous Latency Monitoring with eBPF} \cite{sundberg2023efficient} proposes a framework for continuous one-way delay monitoring with eBPF, demonstrating microsecond precision and low overhead. Although promising, the solution is centered on kernel adjustments and does not address collector modularity.

Similarly, \textit{Analysis and Optimization for Passive One-Way Delay Measurement Tax in Container Networks} \cite{xie2024analysis} investigates passive OWD measurements in container networks, analyzing processing impact and costs. The focus is on dataplane efficiency, without considering software engineering or architectural extensibility.

In distributed-system observability, \textit{ViperProbe} \cite{levin2020viperprobe} and \textit{eBPF-Enhanced Complete Observability Solution for Cloud-Native Microservices} \cite{sharma2024ebpf} apply eBPF to collect metrics, logs, and traces in microservice environments. Both explore data integration for real-time diagnostics but do not present a reusable software model, limiting adaptation to contexts such as SDN or latency-sensitive applications.

Research on integrating eBPF with SDN, such as \textit{Efficient Network Monitoring Applications in the Kernel with eBPF and XDP} \cite{abranches2021efficient} and \textit{Traffic Monitoring Applications Using Programmable Networks and eBPF} \cite{santos2019aplicaccoes}, shows that data-plane instrumentation can be performed directly in the kernel. These works emphasize performance and scalability but do not address standardization of collection software or the use of design patterns.

In SDN architectures, \textit{Protocol-Independent Hardware SDN Router} \cite{pacifico2020sistema} and \textit{Open Chain: Kernel-Based SFC Architecture Using eBPF} \cite{castanho2019cadeia} advance integration between SDN controllers and kernel eBPF functions, but focus on forwarding and filtering rather than latency measurement or modular observability.

On the software engineering side, the literature is scarce regarding formalization of architectural patterns for eBPF-based systems. This work differs by proposing an event-driven collection architecture that incorporates \textit{Observer}, \textit{Strategy}, and \textit{Adapter/Factory}, and by demonstrating that such abstractions maintain performance while reducing evolution effort.

Therefore, this work addresses a practical gap between network programmability and software engineering by proposing a simple and standardized event-driven architecture for continuous latency measurement using eBPF. Unlike prior studies that mainly focus on kernel-level optimizations or dataplane performance, this approach emphasizes modular and reusable software components based on well-known design patterns, while preserving low overhead. As a result, the proposed architecture can be more easily extended and applied in real operational scenarios, including latency-sensitive and safety-critical applications such as connected health systems.

\section{Architecture and Implementation}

This section describes the architecture and implementation of the proposed system, starting with a high-level overview and followed by component-level details.

\subsection{System Overview}

The proposed system is designed to perform continuous latency monitoring with minimal impact on the dataplane while maintaining flexibility for analysis and evolution. The architecture follows a layered design composed of kernel-level instrumentation, user-space agents, and a modular collection and processing layer.

\begin{figure}[t]
    \centering
    \includegraphics[width=0.85\linewidth]{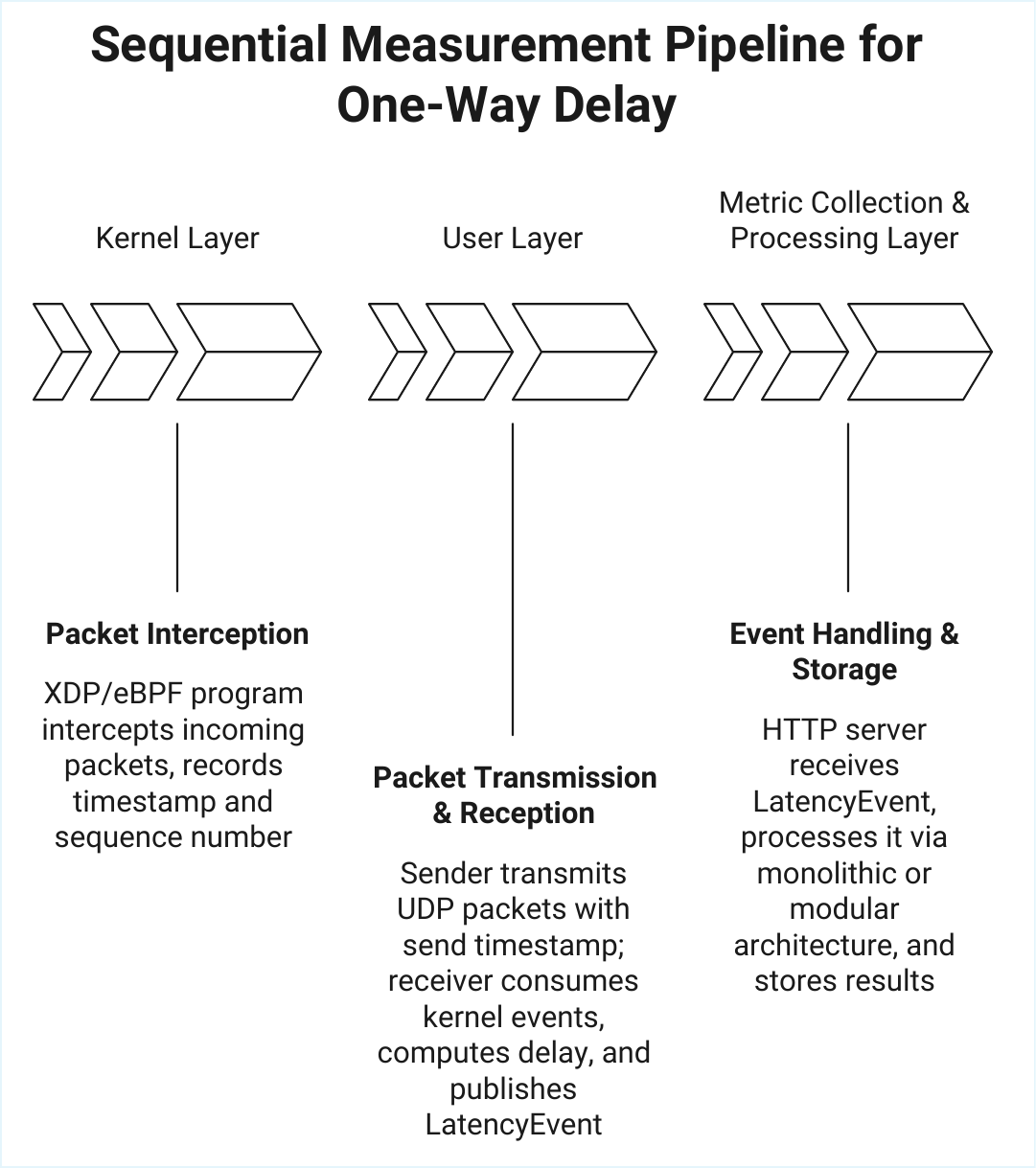}
    \caption{End-to-end pipeline for one-way delay measurement using XDP/eBPF.}
    \label{fig:pipeline}
\end{figure}

Figure~\ref{fig:pipeline} presents the end-to-end pipeline, illustrating how packets are intercepted at the dataplane, transformed into latency events, and processed by modular decision and persistence components.

\subsection{Latency Measurement using eBPF/XDP}

Latency measurements are performed using eBPF/XDP programs attached to the packet reception path. By intercepting packets at the earliest possible point in the networking stack, the system can reach microsecond-level timestamp precision under certain conditions while avoiding interference from higher-layer processing.

Each packet is annotated with sequence identifiers and timestamps, enabling inference of one-way delay (OWD) without relying on clock synchronization between hosts. Since Mininet runs under a shared kernel and time base, OWD can be computed with sufficient precision for latency analysis. This measurement process is continuous and fixed, producing a regular stream of latency samples that serve as input for subsequent decision and reaction mechanisms.

\subsection{Latency Decision and Reaction Strategy}

While latency measurements are continuously collected using eBPF/XDP mechanisms, raw measurements alone are insufficient to guide operational decisions. The system must determine whether observed latency variations correspond to transient fluctuations or sustained performance degradation. To address this, we define an explicit latency-aware decision and reaction strategy, formalized in Algorithm~\ref{alg:latency-decision}.

\begin{small}
\begin{algorithm}[htbp]
\caption{Latency-Aware Decision and Reaction Strategy}
\label{alg:latency-decision}
\begin{algorithmic}[1]
\renewcommand{\algorithmicrequire}{\textbf{Input:}}
\renewcommand{\algorithmicensure}{\textbf{Output:}}

\Require Latency samples stream $L$, window size $W$, thresholds $\theta_{low}$, $\theta_{high}$
\Ensure Monitoring action $A$

\State $L_W \gets$ CollectSamples($L$, $W$)
\Comment{Aggregate samples over a sliding window}

\State $p95 \gets$ Percentile($L_W$, $95$)
\Comment{Estimate tail latency}

\If{$p95 > \theta_{high}$}
    \State $A \gets$ TriggerAlert $\land$ EnableDetailedMonitoring
\ElsIf{$p95 > \theta_{low}$}
    \State $A \gets$ FlagAnomaly $\land$ LogEvent
\Else
    \State $A \gets$ LogMeasurement
\EndIf

\State \textbf{return} $A$

\end{algorithmic}
\end{algorithm}
\end{small}

The proposed strategy aggregates latency samples over sliding windows to reduce sensitivity to short-lived spikes and relies on the p95 statistic to capture tail latency behavior. Two threshold levels are employed to distinguish between moderate and severe degradation, enabling proportional reactions such as logging, alerting, or monitoring escalation. This design preserves dataplane efficiency while helping support timely detection of abnormal latency conditions.

\subsection{Implementation Details}

The proposed architecture was designed to perform continuous one-way delay (OWD) measurements with limited impact on the dataplane and relatively high temporal precision. The system is structured into three complementary layers: (i) Kernel, (ii) User, and (iii) Collector. This organization enables separation of concerns, reduced coupling, and extensions without requiring changes to core measurement logic.

\subsection{Kernel Layer: XDP/eBPF}

In the kernel layer, the eBPF/XDP program intercepts packets upon reception, recording timestamps and sequence numbers at the exact arrival moment. This capture occurs before socket allocation, ensuring microsecond-level precision.

The program operates directly on the RX path, storing information in eBPF maps. Each map entry contains a sequence identifier, arrival timestamp, and send time, enabling inference of one-way delay without relying on inter-host synchronization. Collected events are exported to user space via eBPF maps or \textit{perf events}, ensuring low communication cost and high sampling frequency.

\subsection{User Layer: Sending and Receiving Agents}

The user layer was implemented in Python using the BCC (BPF Compiler Collection) library to communicate with the kernel. Two agents were developed: \textit{sender} and \textit{receiver}.

The \textit{sender} agent transmits UDP packets containing the send timestamp in the payload. The \textit{receiver} agent, running on another Mininet host, captures packet arrival via XDP, computes the one-way delay (OWD), and publishes the result in JSON format to an HTTP collection endpoint.

Publishing occurs continuously, and each latency event is encapsulated in a \texttt{LatencyEvent} structure containing fields \texttt{seq}, \texttt{latency\_us}, and \texttt{timestamp}. This structure standardizes communication across layers and enables the application of design patterns in the collector component.

\subsection{Metric Collection and Processing Layer}

The collection layer was developed in pure Python (\textit{stdlib}), operating as an HTTP server responsible for receiving and processing events sent by the agents. Two collector variants were implemented: (i) \textit{Baseline} and (ii) \textit{Patterns}.

In the \textit{Baseline} version, the HTTP handler is monolithic: each POST request writes directly to a CSV file and, when enabled, applies a fixed threshold to trigger alerts. This version serves as the experimental control.

In the \textit{Patterns} version, the collector adopts a modular architecture structured around three design patterns:

\begin{itemize}
    \item \textbf{Observer:} latency events are published on an \textit{EventBus}, allowing multiple subscribers to react independently.
    \item \textbf{Strategy:} interchangeable analysis policies are defined, such as fixed thresholds or percentile-based sliding window analysis (\textit{P95WindowStrategy}).
    \item \textbf{Adapter/Factory:} persistence and alert sinks (e.g., \textit{CsvSink}, \textit{JsonlSink}, \textit{AlertSink}) are instantiated uniformly without coupling to the collector core.
\end{itemize}

The HTTP handler materializes each received event as a \texttt{LatencyEvent} and publishes it on the bus. Persistence, detection, and alerting logic are delegated to configured modules, keeping the core lightweight, reusable, and extensible.

\subsection{Metrics and Experimental Processing}

Two complementary types of metrics were collected to evaluate both system behavior and architectural impact:

\begin{itemize}
    \item \textbf{One-Way Delay (OWD, $\mu$s):} computed as the difference between send and arrival timestamps obtained directly from the XDP sensor.
    \item \textbf{Evolution Effort:} measured as the number of lines of code added or modified to implement equivalent extensions (e.g., \textit{p95/20} detection and \textit{JSONL} persistence), computed using \texttt{git diff --shortstat}.
\end{itemize}

To ensure experimental hygiene, a warm-up phase was applied by discarding the first five samples of each run. Additionally, outliers were trimmed using the statistical rule $Q3 + 3 \cdot IQR$ to mitigate transient effects of the virtualized environment.

These practices ensured consistency across runs and reliability of the collected metrics, enabling fair comparison between the \textit{Baseline} and \textit{Patterns} implementations. Figure~\ref{fig:pipeline} summarizes the complete measurement pipeline, from kernel-level packet timestamping to modular metric processing and persistence.

\section{Evaluation}

This section presents the experimental setup and evaluation results used to assess the proposed architecture. The experiments focus on performance, measurement accuracy, and modularity-related overhead.

\subsection{Experimental Environment}

Experiments were conducted in an emulated environment to assess the proposed architecture under controlled, reproducible conditions. Mininet enabled simulating network topologies with isolation and precise control of traffic parameters while keeping the same Linux kernel and a shared time base among nodes.

The experimental environment was built using a robust stack of networking and development tools, centered around Mininet for SDN topology emulation and Open vSwitch (OVS) to manage virtual switching. The core logic for the agents and the HTTP collector was developed in Python 3.8+, leveraging the BCC (BPF Compiler Collection) to provide a Python-based interface for interacting with the eBPF subsystem. Additionally, the collector was implemented using standard Python libraries, specifically http.server for network communication and csv and json for efficient data handling and storage.

The topology comprised two hosts (\texttt{h1, h2} and one switch interconnecting them (\texttt{s1}), plus an external host \texttt{r0} running the HTTP collector.  In this network setup, \textit{h1} acts as the UDP packet sender responsible for embedding send timestamps in the payload, while \textit{s1} functions as the virtual switch (OVS) handling link-layer forwarding. On the receiving end, \textit{h2} runs a python script to compute One-Way Delay (OWD) and forward the results to r0, an external host connected to the Mininet network that operates an HTTP collector. The environment ran on a virtual machine with \textit{Ubuntu 22.04 LTS}, kernel \textit{6.8.0-40-generic}, 4 vCPUs, and 4\,GB RAM, ensuring stable performance during measurements.

Experimental traffic consisted of UDP flows sent from \texttt{h1} to \texttt{h2} on port \texttt{5555}, at approximately five packets per second for 20 seconds, totaling around 100 packets per run. The receiver agent  executes the XDP program to record arrival timestamps and publish OWD results in JSON format to the collector’s HTTP endpoint. This process enables continuous measurement and correlation of pipeline behavior under different configurations (Baseline and Patterns).

Before each experiment run, it is performed a warm-up period to stabilize measurements and minimize noise arising from system initialization. The first five samples of each run were discarded, and statistical outlier trimming was applied using $Q3 + 3 \cdot IQR$. This methodology ensured result consistency and comparability across implementation variants, isolating the impact of modularization introduced by the design patterns.

\subsection{Experiments}

The experiments were conducted to assess the behavior of the proposed architecture under controlled conditions, 
by comparing a \textit{Baseline} (monolithic) implementation with a \textit{Patterns} version that incorporates 
Observer, Strategy, and Adapter/Factory abstractions. The evaluation addresses three research questions (RQs): 
(i) qualitative aspects related to observability and system structure (\textbf{RQ1}), 
(ii) the impact on measurement performance and precision (\textbf{RQ2}), and 
(iii) the effort required for architectural evolution (\textbf{RQ3}).
Each experimental run generated approximately 100 OWD measurements, which were processed in real time by the HTTP collector.

Quantitative analysis showed that per-request processing time (\textit{processing\_us}) remained stable, on the order of tens of microseconds, with no significant differences between \textit{Baseline} and \textit{Patterns}.

After discarding the first five samples (warm-up) and trimming with $Q3 + 3 \cdot IQR$, we observed:
\begin{itemize}
    \item Median (\textit{p50}) processing time between 21–22~$\mu$s;
    \item Mean between 26 and 27.5~$\mu$s;
    \item 95th percentile (\textit{p95}) below 40~$\mu$s.
\end{itemize}

These results indicate that collector modularization did not show a clearly perceptible overhead in this environment. Differences between versions were statistically negligible within the expected variability of the virtualized environment.

\begin{figure}[htbp]
\centerline{\includegraphics[width=0.35\textwidth]{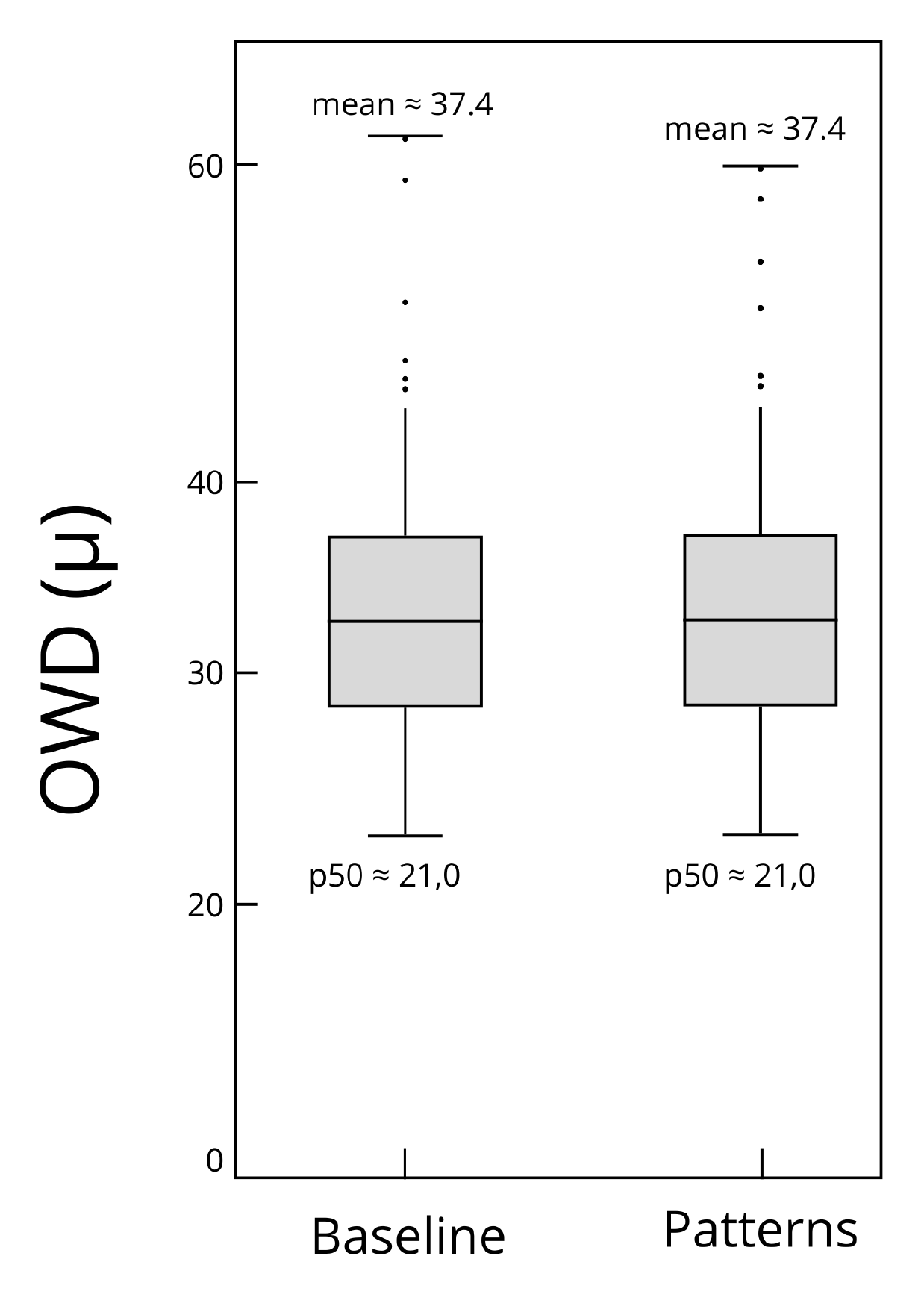}}
\caption{OWD distribution between Baseline and Patterns variants after statistical trimming.}
\label{fig:owd_boxplot}
\end{figure}

The collected OWD statistics presented stable behavior for both implementations. One-way latency measurements (\textit{latency\_us}) were consistent across runs, with minimal variation between \textit{Baseline} and \textit{Patterns}. This demonstrates that the modular pipeline appears to have limited influence on the data plane, preserving collection accuracy even with added observers and dynamic strategies. Figure~\ref{fig:owd_boxplot} illustrates the OWD sample distribution and the similar behavior between the two approaches.

Collector stability suggests that modularization and the use of design patterns do not adversely affect data-path performance, supporting the premise that software abstractions can be introduced without degrading the efficiency of an eBPF/XDP-based system.
To assess architecture extensibility and maintainability, we performed two identical tasks on both collectors: 
(A) implementing a 95th-percentile sliding-window policy (\textit{p95/20}); and 
(B) adding a new data destination (\textit{JSONL sink}). 
The required effort was estimated by measuring the number of lines of code (LOC) changed using 
\texttt{git diff --shortstat}, a commonly used proxy for maintenance and upgrade effort.

Results showed that, for the first evolution (A), the \textit{Patterns} collector required a slightly higher number of lines due to introducing classes and \textit{EventBus} infrastructure. However, subsequent evolutions demanded minimal changes, since the modular architecture allows adding new components without editing the collector core.

From the second task (B) onward, the marginal evolution cost decreased significantly: new \textit{Sinks} or \textit{Strategies} could be added by simply subscribing to the event bus, without changing the HTTP handler. This evidence suggest adherence to the \textit{Open/Closed} principle, enabling extension via composition rather than modification, an essential characteristic in monitoring systems that require rapid adaptation and maintenance.

\subsection{Qualitative Evidence of Observability}

During experiments, the online alert log generated by the \textit{p95/20} strategy  recorded immediate detection of high tails in the first measurement windows. In initial runs, \textit{p95\_us} values near 5.4\,s were observed due to environment warm-up—expected behavior for XDP in \textit{generic} mode. This behavior was mitigated by the warm-up strategy and the statistical trimming applied in the offline analysis. Notably, these adjustments required no changes to the HTTP handler code—only adding \textit{P95WindowStrategy} and \textit{AlertSink}—demonstrating the architecture’s structural flexibility. Direct observation of logs and metrics confirmed that the patterns-based pipeline offered greater readability, traceability, and control over the event flow—desirable traits in continuous monitoring environments.


Table~\ref{tab:resultados} summarizes the main experimental results. The data show that using design patterns brought clear benefits in modularity and extensibility without compromising efficiency or measurement precision. Performance remained equivalent to the monolithic version, and evolution effort decreased as new components were incorporated. These results provide empirical evidence addressing RQ1–RQ3 and suggest that software engineering principles can be applied to eBPF/XDP-based measurement pipelines without sacrificing performance or operational simplicity.

\begin{table}[htbp]
\caption{Summary of metrics obtained in the experiments}
\begin{center}
\begin{tabular}{|l|c|c|}
\hline
\textbf{Metric} & \textbf{Baseline} & \textbf{Patterns} \\
\hline
$p50$ OWD ($\mu$s) & 21.3 & 21.5 \\
$p95$ OWD ($\mu$s) & 38.7 & 39.2 \\
Mean OWD ($\mu$s) & 26.0 & 27.5 \\
Average overhead ($\mu$s) & 22.1 & 23.0 \\
LOC (evolution A) & 47 & 61 \\
LOC (evolution B) & 39 & 17 \\
\hline
\end{tabular}
\label{tab:resultados}
\end{center}
\end{table}

\subsection{Discussion}

The results presented here show that applying software design patterns to eBPF/XDP measurement pipelines is feasible and effective. The modular architecture preserved the same performance as the monolithic collector while improving organization, flexibility, and ease of evolution. This section revisits the RQs in light of the experimental findings.

\subsubsection{RQ1: Architecture and Observability}

The combination of \textit{Observer}, \textit{Strategy}, and \textit{Adapter/Factory} patterns produced a loosely coupled and flexible collector. The core domain, centered on the \texttt{LatencyEvent} and the HTTP handler, remained stable even after adding new strategies and sinks.

Using the \textit{EventBus} as mediator allowed independent reaction to latency events, simplifying testing and maintenance. During experiments, enabling the \textit{p95/20} strategy was sufficient to detect latency variations without modifying the collection logic, demonstrating effective separation of concerns and improved observability.

\subsubsection{RQ2: Performance and Collector Neutrality}

The OWD distributions of the \textit{Baseline} and \textit{Patterns} versions were equivalent. Average processing time remained around 26–27~$\mu$s, with the 95th percentile below 40~$\mu$s. This indicates that introducing abstractions did not compromise data-path efficiency.

Because the collector operates in the control plane and only processes events published by the XDP agent, its overhead is minimal and isolated from the dataplane. Small variations are explained by the virtualized setup, including \textit{xdpgeneric} mode and host OS interference.

These results answer RQ2 by showing that modularization is neutral with respect to performance and suitable for continuous observability systems.

\subsubsection{RQ3: Extensibility and Code Evolution}

Although introducing patterns increases the initial implementation effort, subsequent changes become simpler and more predictable. New components, such as a \textit{JSONL} sink or a percentile-based strategy, were added by implementing new classes and subscribing them to the event bus, without modifying the system core.

This behavior directly addresses RQ3, demonstrating that the architecture supports evolution through extension rather than modification, reducing maintenance effort over time.

\subsection{Practical Implications}

The findings suggest that design patterns are a practical choice for building continuous observability solutions on top of eBPF/XDP. Modularization enables new metrics, policies, and output formats (CSV, JSONL, time series) to be added without destabilizing the system.

More broadly, observability pipelines should be designed as evolvable software systems rather than ad hoc measurement scripts. The results reinforce that modularity and extensibility can be achieved without sacrificing performance, even in latency-sensitive environments.

Collector neutrality with respect to the dataplane confirms that precise measurements can be introduced without interfering with operational traffic, which is essential in real-time and critical applications.

\subsection{Threats to Validity}

The study has limitations related to environment, metrics, and topology. OWD measurements in Mininet rely on a single time base, and outliers were removed using $Q3 + 3 \cdot IQR$. The \texttt{Processing\_us} metric does not capture the full XDP cost but enables fair comparison. The simple topology limits external validity. Small windows may introduce variation, although we used $\geq$20 samples and structured logs.

\section{Conclusion and Future Work}


This work presented a modular architecture for OWD measurement using eBPF, combining kernel-level precision with established software design patterns. We investigated whether abstractions such as \textit{Observer}, \textit{Strategy}, and \textit{Adapter/Factory} could be introduced without compromising performance while improving extensibility and code organization.


Experiments conducted in a Mininet environment show that the modular version maintains processing time and latency behavior comparable to the monolithic baseline, with no observable impact on data-plane efficiency. At the same time, the architecture became easier to evolve, allowing new detection strategies and data sinks to be added with less modification effort. These results indicate that high-performance measurement with eBPF/XDP can coexist with a clean and maintainable software structure.

As future work, we plan to evaluate the architecture in more diverse scenarios, including native XDP deployments and physical environments with time synchronization, as well as extend the pipeline with additional metrics such as jitter and packet loss to further assess its applicability in real-world monitoring systems.

\section*{Acknowledgments}

This work was supported by CNPq (Process 315427/2023-0), FAPERGS (Process 24/2551-0000727-0) and CAPES (Finance Codes 88887.954253/2024-00 and 001). It is also part of the INCT of Intelligent Communications Networks and the Internet of Things (ICoNIoT)(405940/2022-0).

\bibliographystyle{IEEEtran}
\bibliography{meuBibtex}

\end{document}